\documentclass{ws-procs10x7}
\usepackage{balance}

\usepackage{graphicx}
\usepackage{amssymb}
\usepackage{epsfig}

\newcolumntype{d}[1]{D{.}{.}{#1}}

\makeindex
\begin{document}

\title{A High Intensity Linear $e^+e^-$ Collider Facility at low Energy}

\author{A. Sch\"oning}

\address{Institute for Particle Physics, ETH Zurich, CH-8093 Zurich, Switzerland\\$^*$E-mail: schoning@particle.phys.ethz.ch}


\twocolumn[
\maketitle
\abstract{
I discuss a proposal
for a high intensity $e^+e^-$ linear collider
operated at low center of mass energies $\sqrt{s}<5$~GeV with high intensity beams.
Such a facility would provide high statistics samples of (charmed) vector mesons and
would permit searches for LFV with unprecedented
precision in decays of $\tau$ leptons and mesons. 
Implications on the design of the linear accelerator are discussed
together with requirements to achieve luminosities of
$10^{35}$~cm$^{-2}$s$^{-1}$ or more.} 
\keywords{Linear Collider; Tau Factory; Charm Factory.
}]

\section{Introduction}
\begin{figure}[b]
   \begin{picture}(100,175)
     \put(0,0){
       \put(2,-5){\epsfig{file=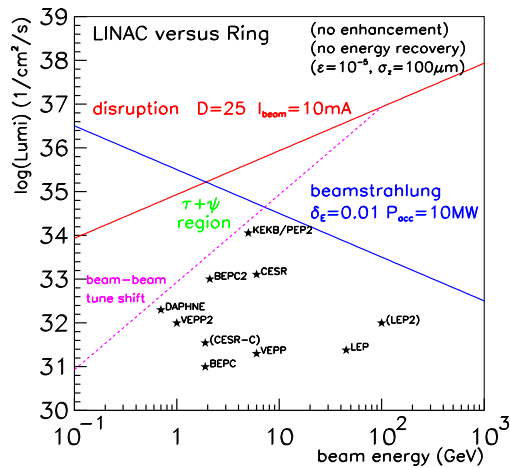,width=0.52\textwidth,height=0.5\textwidth}}
     }
   \end{picture}
\caption{Design luminosities of $e^+e^-$ colliders versus beam energy.
  The broken line shows the upper luminosity limit for ring colliders (eq.~\ref{eq:lumi_ring}).
  The solid lines show the disruption and beamstrahlung limits
  (eq.~\ref{eq:lumi_beamstrahlung}+\ref{eq:lumi_disruption}).
}
\label{fig:collider}
\end{figure}

A tau/charm factory is proposed
delivering luminosities of more than $10^{35}$~cm$^{-2}$s$^{-1}$
and yielding:
\begin{itemize}
\item[$\approx$] $6 \cdot 10^{12}$ $\phi$ mesons p.a.
\item[$\approx$] ~~~\;$10^{12}$ $\psi (2S)$ mesons p.a.
\item[$\approx$] $3 \cdot 10^{10}$ $\psi(3770)$ mesons p.a.
\item[$\approx$] ~~~~$10^{11}$ $\tau$ lepton pairs p.a.
\end{itemize}
The huge luminosity increase compared to existing ring colliders is expected
by using linear collider concepts.
The luminosities of past and existing $e^+e^-$ ring colliders have been 
steadily increased mainly by increasing the beam currents and collision frequencies.
The design luminosities of various $e^+e^-$ ring colliders
are shown in Fig.~\ref{fig:collider} as function of the beam energy.
A main limitation comes from the beam-beam tune shift, which can be written as:
\begin{eqnarray}
L & < & \pi b f \left(\frac{\gamma \Delta Q}{r_e} \right)^2 \;
\frac{\varepsilon_x^*}{\beta_y^*} \nonumber \\
& \propto & \ \gamma^2 \; = \; \frac{E_{\rm beam}^2}{m_e^2} \quad ,
\label{eq:lumi_ring}
\end{eqnarray}
with $b$  the number of colliding bunches, $f$ the rotational frequency,
$\gamma$ the Lorentz boost of the beam particles, 
$ \Delta Q$ the beam-beam tune shift,
$\varepsilon_x^*$ the horizontal emittance
and $\beta_y^*$ the vertical beta function.
The ratio $\varepsilon_x^*/\beta_y^*$ can be
related to parameters of the final focus.
The beam-beam tune shift limit scales with the square of the beam energy and
is the very limiting at low energies.

The luminosity for a linear collider (flat beam) can be written as \cite{ref:tesla_tdr}:
\begin{eqnarray}
L = \frac{1}{4\pi} \frac{P}{E_{\rm beam}} \sqrt{\frac{\delta_E}{0.86 \; r_e^3
    \varepsilon^0_y A}} H_D  \quad ,
\label{eq:lumi_beamstrahlung}
\end{eqnarray}
with $P$ the beam power, $E_{\rm beam}$ the beam energy, $\delta_E$ the relative beamstrahlung, $r_e$ the
electron radius, $\varepsilon_y^0=\gamma \varepsilon_y^*$ the normalized vertical emittance, $A=\sigma_z/\beta^*$
the aspect ratio of the longitudinal beam size over the beta function and
$H_D$ the enhancement factor. The
luminosity is inverse to the beam energy if the beam power and other beam
parameters are kept constant. This limit is in particular important for high beam
energies.

At low energy and high luminosities disruption effects are
important. The disruption parameter depends on the beam
geometry and is for round beams given by:
\begin{eqnarray}
D=\frac{r_e N \sigma_z}{\gamma \sigma_r^2} \quad
\label{eq:disruption_para}
\end{eqnarray}
In the disruption limit the luminosity  can be written as:
\begin{eqnarray}
L = \frac{1}{4\pi r_e m_e} \frac{D \; I \; E_{\rm beam}}{\sigma_z} \quad ,
\label{eq:lumi_disruption}
\end{eqnarray}
with $I$ being the beam current. 
This limit is proportional to the beam energy.

The discussed energy dependencies are shown in fig.~\ref{fig:collider} as lines
for both, linear and ring colliders.
For $I \ge 100\;\mu$A  
a region opens up at low energy, called $\tau$ and $\psi$ 
region, which is accessible by linear colliders but not by ring colliders.
The potentially higher luminosity motivates to use $e^+e^-$ linear
colliders as alternative, not only for highest energies to avoid 
synchrotron radiation, but also for low energies to overcome the fundamental beam-beam
tune shift.

A further advantage of linear colliders is the
flexibility to operate at different beam (center of mass) energies 
using the same machine and same detector. 
The specific luminosities can be significantly higher for a linear collider as dynamic
instabilities, e.g. beam-beam tune shifts, do not play a r$\hat{\rm o}$le. 
Beam disruption might even enhance the luminosity (pinch effect).

\section{General Design Considerations}

Considering the $e^-$ and $e^+$ accelerator as independent 
and assuming round beams,
one finds the following relation for the luminosity:
\begin{eqnarray}
L & = & \frac{1}{2\pi} 
 \left( \frac{I_- I_+}{0.86 \; r_e^3} \right)^\frac{1}{2} 
 \left( \frac{\delta E_- \delta E_+ }{\varepsilon^0_- \varepsilon^0_+ A_- A_+}
 \right)^\frac{1}{4} H_D ~~~~
\label{eq:linear_asym}
\end{eqnarray}
The currents $I_\pm$ are related over the beam energies
$E_i$ to the acceleration power $I_\mp=P_\mp/E_\mp$. For high beam energies the
 main constraint comes from the acceleration power, 
which determines operation costs and is assumed not to exceed 10~MW per
beam. 
However, this limit does not hold when energy recovering techniques are applied,
which are developed worldwide
for various FEL applications~\cite{ref:erl_fel}.

In a classical linear accelerator design, where spent beams are dumped,
a main constraint comes from the requirement of large beam currents $I\gtrsim 100~\mu$A,
 which is a technical challenge 
for the $e^+$ source. For the ILC several concepts have been developed to
provide $e^+$ currents of $50$~$\mu$A in trains and with polarization. 
For a low energy linear collider an order of magnitude higher currents are
desirable to achieve luminosities of $L>10^{35}$~cm$^{-2}$/s. 

Eq.~\ref{eq:linear_asym} suggests that low $e^+$ currents  
can (at least partially) be compensated by high $e^-$ currents.
We can also assume different emittances for electrons and positrons
as well as different energies or beam currents. 
This leads to an asymmetric design of the collider.
Asymmetric energies have the
advantage of introducing a boost, which might
experimentally be
favorable for lifetime tags and oscillation measurements.

 \begin{figure*}[th]
{
   \begin{picture}(500,115)
     \put(0,0){
       \put(-5,-5){\epsfig{file=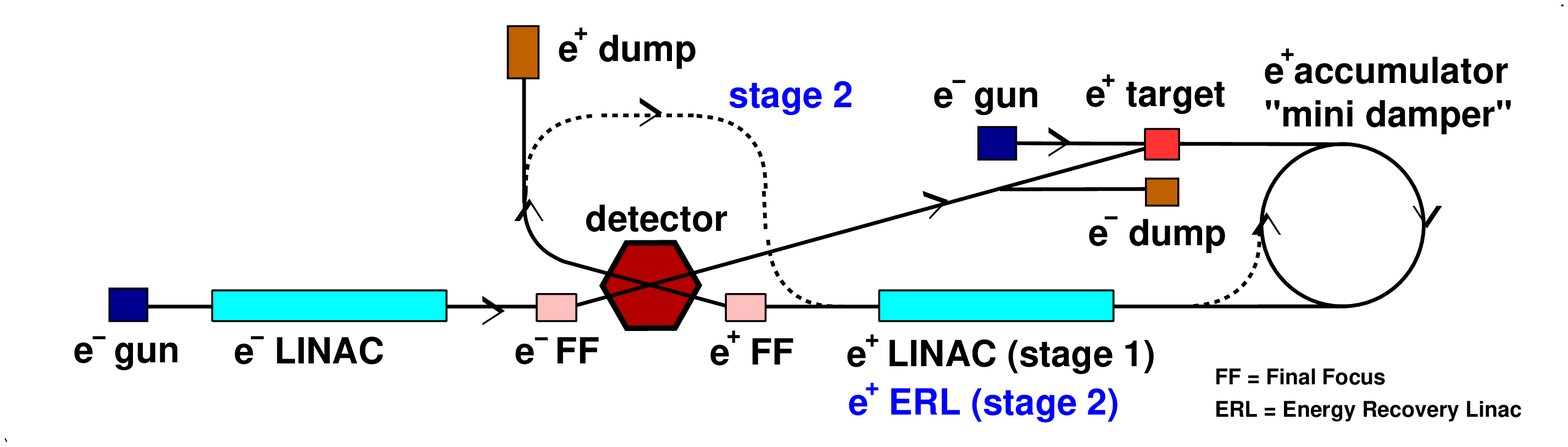,width=1.0\textwidth}}
     }
   \end{picture}
\caption{Schematic design of the proposed Linear Collider at low
  energy. 
}
\label{fig:layout}
}
\end{figure*}

\subsection*{The Asymmetric Collider}
In the following we discuss the case of round beams only, which are produced by standard
 $e^-$ guns, avoid large disruptions in one dimension and allow
for solenoidal final focussing at low energy.
For a given cms energy an asymmetric collider has ten independent machine parameters, 
which determine luminosity and the final focussing system to be
compared to six for a symmetric collider:
{the number of particles in each bunch ($\times 2$)},
{the collision frequency ($\times 1$)},
{the energy ratio of the two beams ($\times 1$)},
{the transverse and longitudinal bunch size ($\times 4$)} and
{the beam emittances ($\times 2$)}.

The luminosity for an asymmetric collider is calculated by:
\begin{eqnarray}
L & = & \frac{f}{2\pi} \frac{N_1 N_2}{(\bar{\sigma}_{r_1}^2 +
  \bar{\sigma}_{r_2}^2)} \nonumber  \quad ,
\label{eq:lumi_linear_round}
\end{eqnarray}
with $\bar{\sigma}_{r_i}={\sigma}_{r_i}/\sqrt{H_D}$ being 
the average transverse beam size during collision
which is determined by the enhancement factor ${H_D}$~\cite{ref:chen_round}.

\subsection*{Beamstrahlung}
Important for the production of narrow resonances is the intrinsic energy spread of the
beam and the energy dispersion resulting from synchrotron radiation 
during collision, called beamstrahlung.
The relative energy loss due to beamstrahlung $\delta_{E}$ should be smaller than
the resonance width. The beamstrahlung is given by~\cite{ref:chen_bs}:
\begin{eqnarray}
\delta_{E} & = & 0.216 \; \frac{N^2 r_e 
  }{\sigma_{r}^2 \sigma_{z}} \; \gamma\; H_{D} 
\quad ,
\label{eq:electron_bs}
\end{eqnarray}
and can be reduced by using flat beams.

By comparing equations \ref{eq:disruption_para} and \ref{eq:electron_bs} we see
that the $\sigma_z$ dependence of disruption and beamstrahlung are
contrarious: 
beamstrahlung is reduced by long bunches and disruption by short bunches. 
Without other boundary conditions maximum luminosity is reached if the disruption and beamstrahlung
limit are fulfilled simultaneously. 
For a tau factory, that is typically the case for a normalized emittance of
$\varepsilon^0=10^{-6}$\;rad\;m and a bunch length of about 50~$\mu$m.

\section{Design Proposal} \label{sec:strategy}
In a classical accelerator design (single collision of beams)
 a high yield $e^+$ source is 
indispensable. 
By using rotating solid targets or liquid metal targets, we assume a
$e^+$ source yielding  $I=100\;\mu$A or more.

Higher colliding $e^+$ currents can be achieved if
the spent $e^+$ beam is captured, possibly damped and re-injected.
The damping ring can be operated at a different (lower) energy than the
colliding beam by decelerating and re-accelerating the beam  using an 
Energy Recovery Linac (ERL)~\footnote{ERLs are considered for the next generation FELs~\cite{ref:erl_fel} to achieve
highest beam intensities in a linear accelerators.}. 
This scheme is interesting because a low energy
damping ring dissipates less synchrotron radiation power~\cite{ref:superb}.
For the proposed linear collider $e^+$ recovery is considered to be more
important than energy recovery in order to increase the current of the
colliding beam. 

Another important issue is the requirement for low emittance beams. 
$e^-$ guns are available or under development which deliver 
peak currents  $I_{\rm peak}>1$~kA 
with emittances of $10^{-6}$ ($10^{-7}$)\;rad\;m in short 
pulses (0.1~ps)~\cite{ref:guns}.
Low emittance positrons can only be produced in damping rings or wigglers but
not from a source. 
A cost effective solution is a 
``mini damping'' ring delivering moderate $e^+$ emittances of
about $\varepsilon=10^{-5}$~rad.m. A small damping ring with short damping 
times $\tau \le 3$-$4 \; \tau_{\rm damp}$ and  $I_{damp}<1$~A 
is considered to be sufficient.

A sketch of the proposed linear collider is shown in
Fig.~\ref{fig:layout}. It shows on the left hand side the $e^-$ machine.
Spent electrons are either dumped or shot on the $e^+$ target
as alternative source to a second $e^-$ gun. 
Produced positrons are accumulated,
damped and accelerated. 
In the first construction stage with single collisions spent positrons
are either dumped or exploited by fixed target experiments.
In the second stage spent positrons are captured and recycled.

 \begin{figure*}
   \begin{picture}(500,130)
     \put(0,-10){
       \put(-5,-5){\epsfig{file=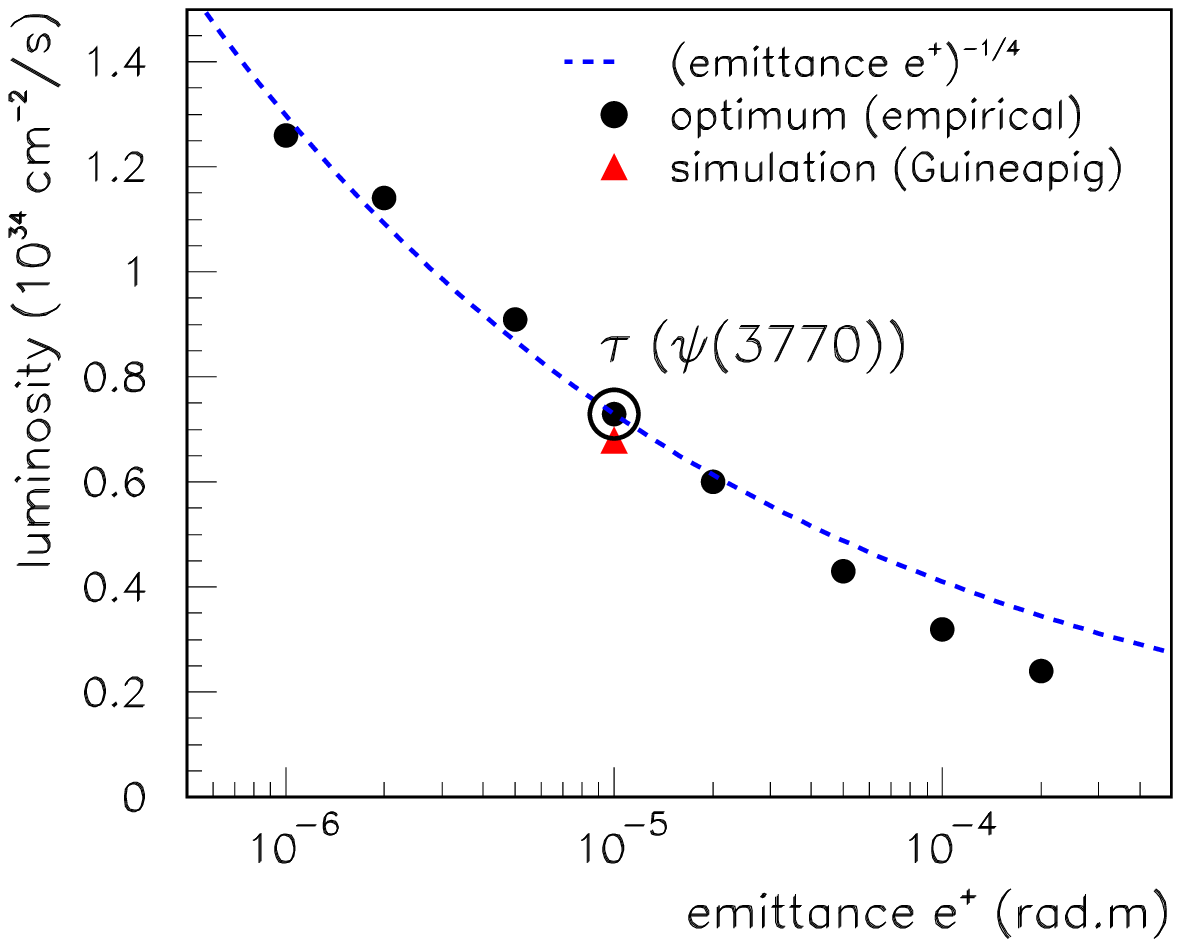,width=0.54\textwidth}}
       \put(205,-5){\epsfig{file=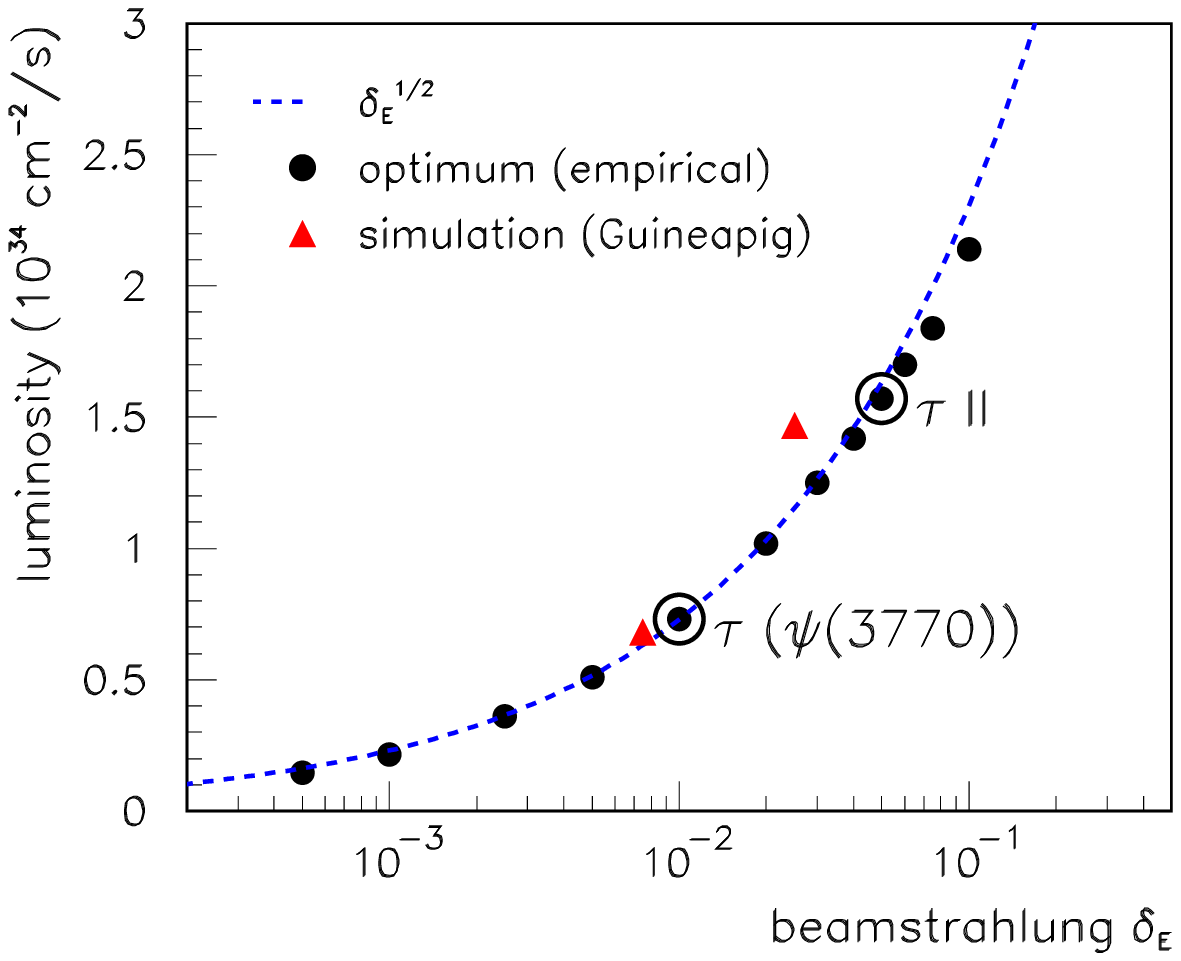,width=0.54\textwidth}}
     }
   \end{picture}
\caption{Luminosity dependence as function of the $e^+$ emittance (left) and
  of the relative beamstrahlung (right) $I(e^+)=100\;\mu$A.}
\label{fig:study_para}
\end{figure*}

\section{Results}
With the above boundary conditions the optimum luminosity is calculated to:
\begin{eqnarray}
L & \approx & \frac{I_+}{2\pi e} \left({\frac{H_D}{0.216 \; r_e^3}} \right)^\frac{1}{2}
\left( \frac{A_- A_+  \delta_{E_-} \delta_{E_+} } {{\varepsilon^0_-
     \varepsilon^0_+ }} \right)^\frac{1}{4}   \quad 
\label{eq:lumi_i2}
\end{eqnarray}
The luminosity dependence as function of different
design parameters was studied in detail for a $\tau$/$\psi(3770)$ factory.
Results obtained by an optimization procedure based on empirical luminosity enhancement
factors are shown in Fig.~\ref{fig:study_para} as
function of the emittance and relative beamstrahlung (black points).  
These results compare well with scaling laws (dashed line)
obtained from eq.~\ref{eq:lumi_i2}.
Two working points for the $\tau$ factories are indicated (open circles).
Results obtained by the beam-beam simulation program Guineapig
\cite{ref:gp} are added (red triangles) which agree  well with the calculations.

Assuming that the colliding $e^+$ current can be increased by a factor of 20
using recycled $e^+$ beams, luminosities of about $10^{34}$~cm$^{-2}$/s
are expected for the $\psi (2S)$ resonance, and  $10^{35}$-$10^{36}$~cm$^{-2}$/s
for the $\phi$ and $\psi(3770)$ resonances and for the $\tau$ pair
production threshold.

\section{Summary}
A high intensity linear collider has been discussed which serves
as high luminosity tau and charm factory providing $e^+e^-$
luminosities of $10^{35}$~cm$^{-2}$s$^{-1}$ or more.
It is concluded that construction of such a low energy collider 
can be started with technology nowadays available. 
Because of the yield limitation of the $e^+$ source
a staged construction is proposed. 
Positron currents can be increased in the second stage 
when more powerful positron targets become available
or by exploiting $e^+$ recovery ($e^+$ recycler) of the spent beam.

\section{Acknowledgments}
I am grateful to L.~Rivkin, K.~Fl\"ottmann and N.~Walker for many helpful and 
stimulating discussions. I thank R.~Eichler for encouraging me to
follow up this interesting topic, and Sylvia for her incessant
support.

\end{document}